\documentclass[usegraphicx,useAMS,usenatbib]{mn2e}
\usepackage{subfigure}

\def\kms{\thinspace\hbox{$\hbox{km}\thinspace\hbox{s}^{-1}$}}
\def\cms{\thinspace\hbox{$\hbox{cm}\thinspace\hbox{s}^{-2}$}}

\def\vsini{\thinspace\hbox{$\hbox{$v$}\thinspace\hbox{sin}$\thinspace$i$}}

\defcitealias{watson_etal_2007b}{Watson et al. (2007b)}
\defcitealias{watson_etal_2007a}{Watson et al. 2007a}

\title[A
  high-latitude star-spot on RU Pegasi]{Roche tomography of
  cataclysmic variables -- V.  A high-latitude star-spot on RU Pegasi}
\author[A. Dunford,  C. A. Watson \&
  R. C. Smith]{A. Dunford$^{1}$\thanks{E-mail: alexander.dunford@gmail.com},
  C. A. Watson$^{2}$\thanks{E-mail: c.a.watson@qub.ac.uk} and Robert
  Connon Smith$^{1}$\thanks{E-mail:
    r.c.smith@sussex.ac.uk}\\ $^{1}$Astronomy Centre, University of Sussex,
  Falmer, Brighton, BN1 9QH, UK\\ $^{2}$Astrophysics Research Centre,
  Queen's University, Belfast, County Antrim, BT7 1NN, UK}
\begin{document}

\date{Accepted ...........  Received ........; in original form
  2011 August 19}

\pagerange{\pageref{firstpage}--\pageref{lastpage}} \pubyear{2012}

\maketitle

\label{firstpage}

\begin{abstract}
We present Roche tomograms of the secondary star in the dwarf nova
system RU Pegasi derived from blue and red arm ISIS data taken on the
4.2-m William Herschel Telescope. We have applied the {\it{entropy
    landscape}} technique to determine the system parameters and
obtained component masses of $M_{1}=1.06$\,M$_{\odot}$,
$M_{2}=0.96$\,M$_{\odot}$, an orbital inclination angle of $i =
43^{\circ}$, and an optimal systemic velocity of $\gamma =
7\kms$. These are in good agreement with previously published values.

Our Roche tomograms of the secondary star show prominent irradiation
of the inner Lagrangian point due to illumination by the disc and/or
bright spot, which may have been enhanced as RU Peg was in outburst at
the time of our observations. We find that this irradiation pattern is
axi-symmetric and confined to regions of the star which have a direct
view of the accretion regions.  This is in contrast to previous
attempts to map RU Peg which suggested that the irradiation pattern
was non-symmetric and extended beyond the terminator.

We also detect additional inhomogeneities in the surface distribution
of stellar atomic absorption that we ascribe to the presence of a
large star-spot. This spot is centred at a latitude of
$\sim$82$^{\circ}$ and covers approximately 4 per cent of the total
surface area of the secondary. In keeping with the high latitude spots
mapped on the cataclysmic variables AE Aqr and BV Cen, the spot on RU
Peg also appears slightly shifted towards the trailing hemisphere of
the star. Finally, we speculate that early mapping attempts which
indicated non-symmetric irradiation patterns which extended beyond
the terminator of CV donors could possibly be explained by a superposition
of symmetric heating and a large spot.

\end{abstract}


\section{Introduction}

Cataclysmic variables (CVs -- see \citet{warner_book} for an excellent
review) are a class of interacting binary characterised by a white
dwarf star, known as the primary, accreting material via an accretion
disc or column from the secondary -- typically a late-type
main-sequence dwarf star. CVs are often studied in order to gain
insight into the processes of astrophysical accretion, but it is the
secondary star that holds the key to unlocking our understanding of
the creation and evolution of these systems and their corresponding
behaviour. In the canonical scenario for the evolution of CVs, for
example, the magnetic field of the secondary star is believed to play
a crucial role in draining away angular momentum from the system via a
process known as {\it{magnetic braking}} (see, e.g.,
\citet{kraft_1967}, \citet{mestel_1968} and \citet{spruit_1983}). This
is  thought to be responsible for maintaining mass transfer and for
causing long period CVs to evolve to shorter orbital periods. At orbital
periods of about 3 hours and stellar masses of around
$0.25$\,M$_{\odot}$, the secondary star is thought to become fully
convective.  Under the canonical theory of CV evolution, this
transition to a fully convective interior switches off the stellar
dynamo and consequently mass transfer is quenched. Contact with the
inner Lagrangian point is re-established at shorter periods through
removal of angular momentum via gravitational radiation. This explains
the so-called {\it{period gap}} in the period distribution of the
total CV population. Despite the acceptance of this model, there is
little or no direct observational evidence for such an abrupt change
in the angular momentum loss mechanism and it is still unclear whether
magnetic braking is really the dominant loss mechanism above the
period gap. Magnetic activity cycles have also been cited as
possible causes for a number of other observed CV phenomena, such as
variations in orbital periods, luminosity, mean outburst durations and
outburst shapes (e.g. \citet{applegate_1992}, \citet{richman_1994} and
\citet{Ak_2001}). Given the importance of understanding the underlying
magnetic properties of these stars, it is therefore highly desirable
to try to obtain observational evidence for such fields, in the form
of the sizes, distributions and lifetimes of star-spots.

In addition, since CV secondaries are among the fastest rotating set
of stars known, we may begin to use them as natural laboratories to
provide tests of stellar dynamo theory, since they should display
strong dynamo action. We may also begin to ask what effect tidal
forces have on the emergence of magnetic flux tubes. It has been
suggested that tidal forces may suppress differential rotation
\citep{scharlemann_1982} and force star-spots to form at preferred
longitudes \citep{holzwarth_2003}. Since CV secondaries are heavily
tidally distorted, it seems clear that these systems are perfect for
probing these effects.

\begin{figure*}
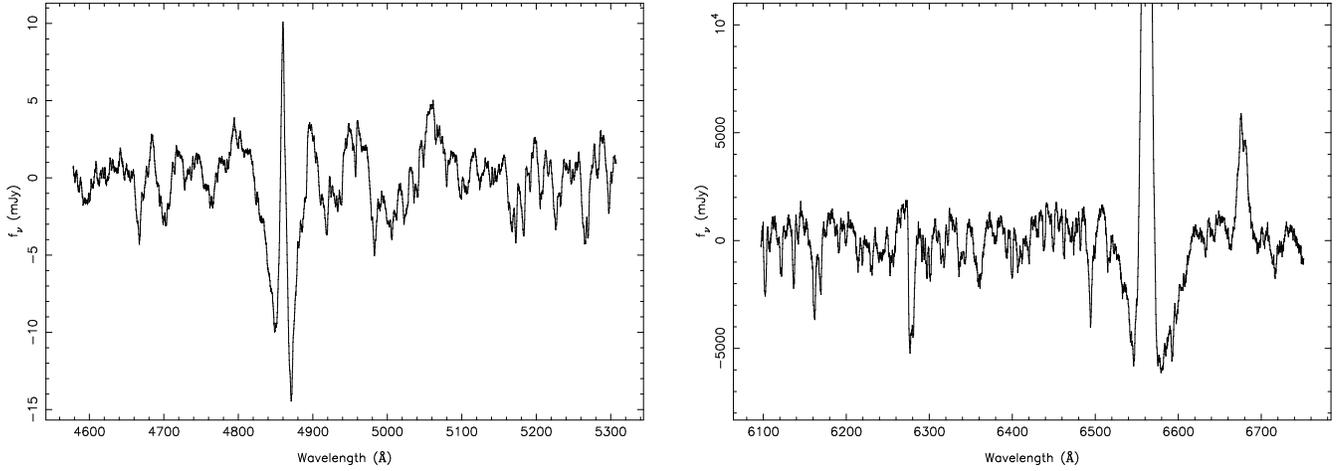

\centering {
    \label{average_rupeg}
    \includegraphics[scale=0.35,angle=270]{average_rupeg.eps} }
\hspace{0.3cm} {
    \label{average_rupeg_red}
    \includegraphics[scale=0.35,angle=270]{average_rupeg_red.ps} }
\caption{{\it{Left}} -- WHT+ISIS average blue arm spectrum of RU
  Pegasi. {\it{Right}} -- WHT+ISIS average red arm spectrum of RU
  Pegasi. In both cases the orbital motion of the secondary star has
  been removed prior to co-adding, and the continuum has been
  subtracted.}
\label{spectra}
\end{figure*}

However, despite the clear importance of magnetic behaviour in the
evolution of CVs and the potential gains to be had in our
understanding of the so-called solar-stellar connection, until
recently there has been little work undertaken in this
field. \citet{webb_2002} used TiO bands to determine a spot coverage
factor of 22 per cent for the secondary in SS Cyg, although their work
did not actually map the surface. Recently, however, the technique of
Roche tomography has begun to allow the mapping of the surface of
these stars, and results presented in papers III and IV of this series
have reported the first detection of star-spots on a CV secondary (AE
Aqr -- \citealt{watson_etal_2006}) followed by a similar mapping
campaign for BV Cen \citepalias{watson_etal_2007a}. Both maps showed spots
at almost all latitudes, as well as large near-polar spots similar to
those found in Doppler imaging studies of single stars. In this work,
we report on attempts to apply this technique to the dwarf nova RU Peg
for the first time. RU Peg is a relatively bright ($m_{\rm v}$=13.1
rising to 9.5 during outburst) U Geminorum type dwarf nova with an
orbital period of 9 hours.

\section{Observations and data reduction}

\subsection{Observations}
\label{sec:obs}

Observations of RU Pegasi were made with the 4.2-m William Herschel
Telescope (WHT) at the Roque de los Muchachos Observatory on the
island of La Palma over the nights of 20--23 July 2004. A journal of
these observations can be found in Table {\ref{journal}}. The ISIS
double beam spectrograph, which makes use of a dichroic mirror to
split the incoming light into blue and red channels, was used to
obtain simultaneous blue and red spectra. The central wavelengths for
the blue and red spectra were selected to be 4950\AA \, and 6400\AA \,
respectively, yielding unvignetted ranges of 4550--5350\AA \, in the
blue and 6070--6700\AA \, in the red. Both arms of ISIS made use of
the R1200 gratings giving a dispersion of about 0.23\AA/pixel. A
two-pixel resolution of about 28\kms \, in the blue and 22\kms \, in
the red was obtained for a slit width of 1 arcsec. Exposure times were
selected in such a way as to balance the requirement for maximum
signal while remaining short enough to avoid the effects of velocity
smearing. Observations of the flux standard star SP\,2148+248 were
taken at or close to the zenith, in order to correct for the
instrumental response and to remove any telluric absorption features
in the red data.

Since we require good relative flux calibration, we placed the nearby
comparison star RU Peg Ref 2 (\citealt{harrison04}) on the slit to
allow for slit-loss corrections and any variations in sky transparency.
In addition to a number of observations of K and M
spectral-type templates, arc spectra were taken before and after each
exposure (or multiple run of six ten minute exposures in the case of
RU Peg) to enable the spectra to be wavelength calibrated and to
correct for any flexing in the instrument as the telescope tracks the
target. A set of tungsten flat fields and bias frames was also taken
at the beginning and end of each night. The weather during the
observing run varied from clear to heavy rain and fog. In effect, this
resulted in the loss of one night out of four. Mechanical breakdowns
also restricted the amount of observing we were able to
accomplish. Amateur $V$ band photometric observations (obtained by
Roger Pickard and Gary Poyner of the British Astronomical Association
Variable Star Section (BAA VSS), whom we gratefully acknowledge)
showed that the system was not in quiescence during our
observations. Rather, it was on the rise to peak outburst
luminosity. The average blue and red arm spectra can be seen in
Fig. {\ref{spectra}}.

\subsection{Data reduction}

The raw images were 
bias-subtracted using a median bias value obtained from the overscan
region. Pixel-to-pixel variations were removed by creating a master
balance frame, generated from a set of tungsten flats, and dividing
each raw frame through by the master balance frame.  
The spectra were finally
sky subtracted and optimally extracted (\citealt{horne_1986}) using
the software package \texttt{PAMELA}. The Copper Argon/Copper Neon
(CuAr+CuNe) arc spectra were also optimally extracted, 
from the same region of the chip that
the object spectrum occupied. These were fitted with a fourth order
polynomial giving an rms scatter of better than 0.001\AA. The object
spectra were then placed on this wavelength scale. The data were then
flux calibrated using spectra of the featureless white dwarf star
SP\,2148+286.

Roche tomography also requires that we correct for slit-losses and
variable sky transmission. This is due to the rapidly varying and
unknown spectral contribution of the accretion disc. Because ISIS is
not a cross-dispersed spectrograph, and so is capable of long-slit
spectroscopy, we were able to place a non-varying comparison star (RU
Peg Ref 2; \citealt{harrison04}) on the slit and take wide and narrow
slit exposures. We then multiplied each corresponding CV spectrum by
the ratio of the wide slit comparison spectrum to the narrow slit
comparison spectrum. We should note that since the slit was not set to
follow the parallactic angle, this is at the expense of some small
effects arising from differential refraction. Since we treat both the
red and blue arms independently during the Roche tomography
reconstructions later, this reduces our exposure to such effects.

\begin{table*}
\begin{center}
\begin{tabular}{|c|c|c|c|c|c|l|}
\hline Object & Date & UT Start & UT End & $T_{\rm exp}$ (s) & No. of
spectra & Comments \\ \hline
BD+03 3568 & 2004 July-20 & 23:08 & 23:23 & 300 & 2 & G8V template
\\ BD+05 3625 & & 23:25 & 23:35 & 600 & 1 & K0V template \\ GAT 746 &
& 23:52 & 00:02 & 600 & 1 & K1V template \\ RU Peg & & 00:22 & 02:11 &
600 & 13 & CV target \\ SP 2148+286 & & 02:42 & 02:52 & 600 & 1 &
Spectrophotometric standard \\ SP 2148+286 & & 02:53 & 03:03 & 600 & 1
& Spectrophotometric standard wide slit\\ RU Peg & & 03:15 & 05:32 &
600 & 13 & CV target \\
HD 154363 & 2004 July-21 & 21:53 & 21:55 & 120 & 1 & K5V template
\\ HD 181196 & & 22:03 & 22:05 & 180 & 1 & K5V template \\ HD 332601 &
& 22:09 & 22:14 & 600 & 1 & K5V template \\ HD 160964 & & 22:24 &
22:28 & 240 & 1 & K4V template \\ SP 2148+286 & & 00:34 & 00:44 & 600
& 1 & Spectrophotometric standard \\ SP 2148+286 & & 00:45 & 01:06 &
600 & 2 & Spectrophotometric standard wide slit \\ RU Peg & & 01:28 &
05:41 & 600 & 45 & CV target \\
BD+00 3459 & 2004 July-22 & 22:49 & 22:50 & 60 & 1 & K3V template
\\ RU Peg & & 02:35 & 04:37 & 600 & 22 & CV target \\ BD+30 3627 &  &
04:47 & 04:52 & 300 & 1 & K2V template \\ SP 2148+286 & & 04:56 &
05:06 & 600 & 1 & Spectrophotometric standard \\ HD 333388 & & 05:14 &
05:16 & 150 & 1 & K8V template \\
BD+01 3236 & 2004 July-23 & 21:39 & 21:47 & 450 & 1 & K7V template
\\ GJ 9606 &  & 22:10 & 22:20 & 600 & 1 & M0V template \\ GJ 9645 &  &
22:27 & 22:35 & 450 & 1 & K6V template \\ SP 2148+286 & & 22:41 &
23:15 & 600 & 3 & Spectrophotometric standard \\ RU Peg & & 02:26 &
03:48 & 600 & 13 & CV target \\
\hline
\end{tabular}
\caption{Journal of observations for the July 2004 WHT/ISIS run. The
  first column gives the object name, followed by the date and
  start/end times of the observations. The final three columns
  indicate the exposure times (for the blue arm of ISIS), the number
  of spectra obtained and an indication of the type of spectra taken.}

\label{journal}
\end{center}
\end{table*}

\section{The secondary star}

\label{secondary_star}

\subsection{Projected rotational velocity}\label{prv}

In order for accurate orbital phases to be calculated for our spectra,
a new orbital ephemeris was required (see section
{\ref{rv}}). However, in order to provide the most accurate
cross-correlation between our target spectra and the template, we also
need an estimate for the mean line width of the target spectrum. To be
able to measure the mean projected equatorial rotational velocity of
the secondary, $\vsini$, we require an average spectrum for maximum
signal to noise. Simply creating a mean spectrum at this point would
be futile, as the orbital motion of the secondary would effectively
smear out the absorption line profiles due to the Doppler shifts
throughout the orbit. We therefore subtracted the radial velocity of
each spectrum individually, having determined their velocities through
cross-correlation with a high signal-to-noise spectral-type template
star. The spectra are, of course, composites of all the components in
the CV. To ensure the secondary star line strengths were preserved
across the region of interest, the spectra were first rebinned onto
the same velocity scale using sinc-function interpolation and then
normalised by dividing through by a constant. Finally, the spectral
continua were fitted with a third-order spline which was then
subtracted. An identical procedure was applied to the spectral type
template with the added complication that the spectral type standard
was corrected for its systemic velocity.

The majority of our template stars do not have prior systemic velocity
measurements and so we used Least Squares Deconvolution (LSD -- see
section {\ref{lsd}}) to calculate a mean absorption line profile for
each spectrum. This mean profile was then fitted with a gaussian and
the location of the peak of the fit was taken as the template's radial
velocity. Table 2 lists the measured radial velocities of our spectral
type templates. A previous estimate of the radial velocity of our K5V
template GL 653 can be found in \citet{evans_1967} who reported a
value of $33.6\pm2\kms$. This is within 1.5\,$\sigma$ of our LSD
derived estimate of $30.6\kms$.

\begin{figure}
\begin{center}
\includegraphics*[scale=0.35, angle=270]{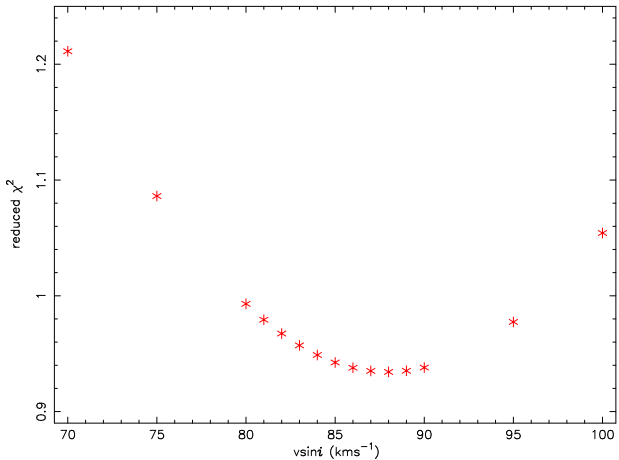}
\end{center}
\caption{$\chi^{2}$ versus $\vsini$ obtained through cross correlation
  with a K5V template. The minimum is located at a projected
  rotational velocity of 88\kms.  Note (see text) that the broadening 
increment of 5\kms\ has been reduced to 1\kms\ around the minimum to
allow its position to be better determined.}

\label{k5vchi}
\end{figure}

A first iteration on the radial velocity curve was obtained by
cross-correlating the RU Peg spectra with a K5V template that had been
rotationally broadened by an arbitrary amount, $100\kms$ in this
instance. We used the red arm data, specifically wavelengths between
6200\AA \, \& 6500\AA, since this range contains a number of
relatively strong stellar absorption lines and also some known
temperature and gravity sensitive lines for K type stars
\citep{strassmeier_fekel_1990}. This first pass made it possible to
subtract approximate radial velocities of the spectra and create an
average RU Peg spectrum.

The template spectrum was then artificially broadened by $5\kms$
increments, multiplied by some constant and then optimally subtracted
from the average RU Peg spectrum. In this process, the fraction of the
template that best removes the secondary star lines is calculated
using an optimisation technique that minimises the scatter between the
template and the target in a process known as {\it optimal
subtraction} (a term first used by \citet{dhillon_1993} and widely used 
since in the context of secondary stars in CVs). 
Fig. \ref{k5vchi} shows the results of this method for
the K5V template, where we have decreased the broadening increment to
$1\kms$ around the curve's minimum. The optimum value of \vsini \, is
88\kms.

To test whether our choice of spectral type template introduces
systematic errors, or indeed if there is any trend in the measured
value of $\vsini$ with spectral type, the procedure has been repeated
for all our spectral type templates. We find that the choice of
spectral type does not significantly affect the resulting optimal
value for $\vsini$ -- the standard error on the mean of our sample is
only $0.5\kms$. In each application of artificial broadening, we have
assumed a simple linear limb darkening law with a coefficient of 0.5,

\begin{equation}
\frac{I(\mu)}{I(0)}=1-a(1-\mu)
\end{equation}

\noindent where $\mu=\cos\delta$ ($\delta$ being the angle between the
line of sight and the emerging intensity), $I(0)$ is the specific
intensity at the centre of the stellar disc and $a$ is the linear limb
darkening coefficient \citep[see][]{claret_1998}. Again, to test for
the possibility of introducing systematic errors, the entire procedure
has been run using a range of values from 0 to 1 for the limb
darkening coefficient. The optimum values of $\vsini$ did not differ
from each other by more than $3\kms$. We therefore adopt $v \sin i$ =
88 $\pm$ 3\,km\,s$^{-1}$.

\begin{figure}
\begin{center}
\includegraphics*[scale=0.35,angle=270]{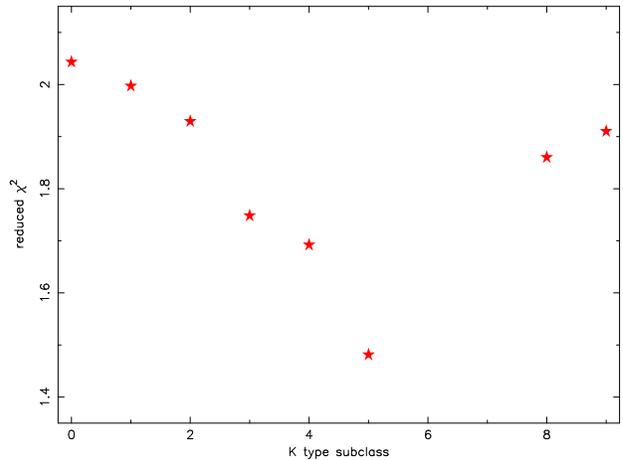}
\end{center}

\caption{RU Peg secondary star spectral type. The $x$-axis refers to
  the subclass of K dwarf template used in the optimal subtraction
  procedure described in section {\ref{prv}}.}

\label{sptype}
\end{figure}

The spectral type of the secondary star can also be estimated using
optimal subtraction, given a full set of spectral type
templates. Fig. {\ref{sptype}} shows the reduced $\chi^{2}$ against K
dwarf subtype which displays a minimum at K5V. Again, we tested the
effect limb darkening has on the outcome. We found no change in the
optimum spectral type while varying the linear limb darkening
coefficient. Our estimated spectral-type compares well with the
previous estimated spectral type of K3V adopted by
\citet{friend_et_al_1990a} following the work of \citet{wade_1982}.

\subsection{Radial velocity curve and ephemeris}\label{rv}

The most recent orbital ephemeris for RU Peg was derived from
secondary star radial velocity variations by
\citet{friend_et_al_1990a}. Given the long gap between their
observations and the ones reported here, a large error may have
accumulated in the orbital zero point. We have derived a new ephemeris
by cross-correlating our red arm RU Peg data with the best-fitting K5V
spectral type template GL 653 to produce a series of cross-correlation
functions (CCFs). The RU Peg and the GL 653 spectra are treated in the
same way as described in section {\ref{prv}}, with the template being
artificially broadened by the optimum amount found through optimal
subtraction, in this case $88\kms$. In addition, the radial velocity
of the K5V template has been removed. A cross correlation function was
generated for each RU Peg spectrum, and the radial velocity curve
obtained by fitting a sinusoid through the CCF peaks. The following
ephemeris:

\begin{equation}
T_{0}=245 3207.533 76 \pm 0.00009,
\end{equation}

\noindent was obtained where the orbital period has been fixed at
$0.3746$ days \citep{stover_1981}, and $T_{0}$ is the time of inferior
conjunction of the secondary star, in the heliocentric reference frame. 
The quoted uncertainty is statistical, with no allowance for any possible
systematic effects.
This new zero point has been used
in all subsequent analysis. The robustness of this method of
generating a radial velocity curve through cross-correlation against
an incorrect template choice has been tested. A series of CCFs was
produced using the full range of K dwarf templates at our disposal,
and it was found that, providing the templates were correctly
broadened, there was no significant change in the value of the
semi-amplitude or systemic velocity, and therefore the
zero-crossing.

\begin{figure}
\begin{center}
\includegraphics*[scale=0.35, angle=270]{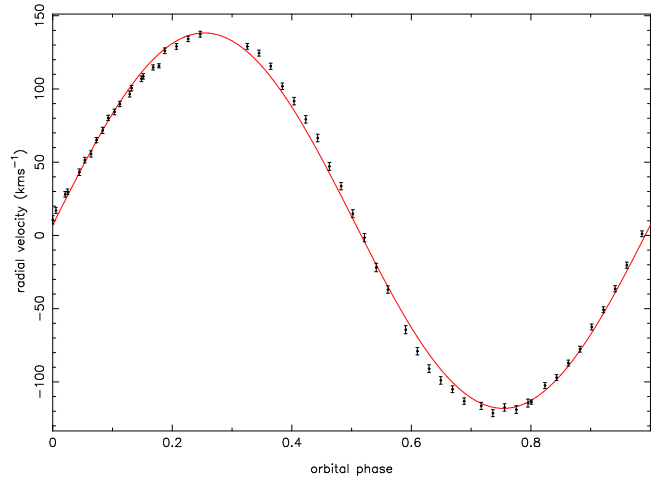}
\end{center}
\caption{RU Peg (ISIS red arm) radial velocity data (filled circles)
  and circular orbital fit.}
\label{rvcurve}
\end{figure}

In Fig. {\ref{rvcurve}}, the radial velocity points
have been assigned an orbital phase based on the new ephemeris, and
fitted with a circular fit of the form,

\begin{equation}
V\left(\phi\right)=\gamma+K_{2}\sin{\frac{2\pi\left(t-t_{0}\right)}{P_{\rm
      orb}}},
\end{equation}

\noindent where $K_{2}$ is the semi-amplitude of the secondary star,
$\gamma$ is the systemic velocity, $V\left(\phi\right)$ is the radial
velocity at a phase $\phi$, $t$ is the time of observation, $t_{0}$ is
the zero crossing and $P_{\rm orb}$ is the orbital period. From this
fit we obtain a value for the radial velocity semi-amplitude of
$128.2\pm0.6\kms$. This compares with a value of $111\pm8\kms$ found
in \citet{kiplinger_1979}, $121\pm2\kms$ from \citet{stover_1981} and
$121\pm2\kms$ derived in \citet{friend_et_al_1990a}. We should add at
this point that there is a clear systematic trend in the residuals of
the circular fit of Fig.~\ref{rvcurve}.  This is most likely the
result of irradiation, making such conventional studies susceptible to
systematic errors since the centre-of-light of the secondary no longer
represents its centre-of-mass.  Our value for the systemic velocity of
$10\pm0.5\kms$ compares well with \citet{friend_et_al_1990a} who
report a value of $13\pm2\kms$, and Stover's value of $5\pm4\kms$. The
analysis of \citet{kiplinger_1979} yielded a slightly more discrepant
value of $-1\pm5\kms$ using the radial velocity data obtained in
\citet{kraft_1962}, though the value changes to $18\pm6\kms$ when
measuring $\gamma$ from a disc emission line radial velocity curve.

Additional motivation for measuring the projected rotational velocity
and $K_{2}$ is that one can determine a value for the mass ratio and
in turn, the semi-amplitude of the primary star. From
\citet{horne_etal_1986}, the predicted rotational velocity for a
Roche-lobe filling star is given by:

\begin{equation}
\vsini=\frac{R_{\rm L}}{a}K_{2}\left(1+q\right).
\end{equation}

\noindent Combining this with the approximation of \cite{smith_dhillon_1998} for
the volume-averaged Roche lobe radius (assumed equal to the radius of
the secondary star) gives the following relation:

\begin{equation}\label{forq}
\vsini=0.47K_{2}\,q^{1/3}\left( 1+q \right)^{2/3}.
\end{equation}

\noindent Rearranging for $q$, the mass ratio (defined as
$M_{2}/M_{1}$), and inserting our values for $\vsini$ and $K_{2}$
derived earlier, equation ({\ref{forq}}) can be solved to obtain
$q=0.88\pm0.02$. This is higher than the values of 0.78 and 0.79 found
in \citet{stover_1981} and \citet{kiplinger_1979}, both obtained
directly through measurements of both $K_{1}$ and $K_{2}$. Given the
relatively poor quality of their emission line radial velocity curves,
it may be argued that the value derived here from accurate rotational
velocity measurements is somewhat more reliable, although our value
for $K_{2}$ has not fully taken into account the residuals in the
fit. \citet{friend_et_al_1990a} derive a value of $q=0.82\pm0.15$ from
their rotational broadening measurements that is slightly closer to
ours. However, the value presented here is considered an improvement
given the higher resolution of our data. Our mass ratio and $K_{2}$
imply a primary star radial velocity semi-amplitude
$K_{1}=112.6\pm0.25\kms$. \citet{kiplinger_1979} reports a value of
$88\pm10$\kms, significantly lower than our value, from direct
measurement  of the H${\alpha}$ emission wings; although the fit is
reasonable, the reported rms error is $26$\kms. \citet{stover_1981},
using a similar method for measuring the emission wing velocities,
finds a value of $94\pm3$\kms \, fitting a circular orbit to the
radial velocity data (with an rms error of $12$\kms). We note here
that no attempt has been made to directly measure the emission line
radial velocities in this work due to the complicated nature of the
emission line profiles. Since the system is in outburst, we see that
H${\beta}$ and H${\alpha}$ have two anti-phased components -- one from
the disc and one from the irradiated surface of the secondary.

\begin{table}
\begin{center}
\begin{tabular}{|c|c|c|}
\hline Spectral Type & Name & Radial Velocity (\kms) \\ \hline G8V &
BD+03 3568 & -0.3 \\ K0V &  BD+05 3625 & -31.9 \\ K1V & GAT 746 &
-50.0 \\ K2V & BD+30 3627 & -57.3 \\ K3V & BD+00 3459 & -5.3 \\ K4V &
HD 160964 & -20.8 \\ K5V & GL 653 & 30.6 \\ K8V & HD 333388 & -7.1
\\ \hline
\end{tabular}

\caption[Heliocentric radial velocities of spectral type templates
  (WHT data)]{Heliocentric radial velocities of spectral type
  templates measured via a Gaussian fit to the mean profile.}

\end{center}
\label{temp_rv}
\end{table}

\section{Least squares deconvolution}\label{lsd}

While single absorption-line studies of CV donor stars using
intermediate resolution spectrographs on 4-m class telescopes have
been successful in mapping large-scale irradiation patterns on their
surfaces \citep[e.g.][]{watson_et_al_2003}, they have not been
successful at mapping star-spots because of the limited resolution and
signal-to-noise of such data. \citet{watson_etal_2006}, however,
applied the technique of Least Squares Deconvolution (LSD --
\citealt{donati_1997}) to produce a mean profile by combining several
thousand absorption lines from the secondary star in AE Aqr observed
with the UES echelle spectrograph on the WHT.  This achieved the
required signal-to-noise to observe the effects of star-spots on the
line profile.

LSD requires the positions and strengths of the atomic absorption
lines present in the stellar spectrum to be known. In this work, we
use an atomic line list generated by the Vienna Atomic Line Database
(VALD, \citet{kupka_1999, kupka_2000}). VALD offers a simple online
interface for the selection of the required data; all that is required
are values for the stellar effective temperature and surface gravity,
the relevant wavelength range, micro-turbulence (due to small scale
motions of convective cells at the stellar surface, taken here to be
$\approx$1\kms) and a detection limit for the strengths of the
normalised lines. The detection limit in Doppler imaging applications
is often taken to be about 0.4 (this is the lower limit to the central
line depth, ignoring line blending and expressed as a fraction of the
continuum). Since we are dealing with intermediate resolution data
rather than echelle data, we  have used a more conservative 0.5 (this
may still seem optimistic; however tests described below indicate that
the input line list has little effect on the shape of the mean
profile). The spectral type of RU Peg has been estimated in this work
as K5V -- the closest model available in the VALD database corresponds
to a stellar atmosphere with an effective temperature, $T_{\rm eff}$,
of $4600$K and $\log{(g/\cms)}=4.6$.

The composite nature of the CV spectrum complicates the situation with
regard to treating the spectral continuum. There is a varying and
unknown contribution from the accretion regions, as well as a changing
continuum slope due to any time variability and the changing sky
projection of the accretion disc and hot spot. This means that the
creation of a master continuum fit \citep[e.g.][]{acc_1994} is
impossible and we are forced to remove the continuum on a spectrum by
spectrum basis. Simply normalising the spectra is also inappropriate,
as the varying secondary star contribution would result in the
relative line strength changing from one spectrum to the next. We
therefore continuum subtracted the data (after flux calibration and
slit-loss correction) by subtracting a smoothly varying fit for each
spectrum. This fit is created by placing spline knots at positions
that are relatively free of absorption lines, while simultaneously ignoring
strong disc emission lines.

This method of fitting the continuum may appear rather subjective,
since it is often difficult to decide exactly how the continuum varies
across a spectrum. Tests were carried out to determine the effects of
incorrect continuum fits on the final mean profile by constructing two
continuum fits to the same spectrum -- the first being our best guess
at the true continuum and the second was a fit which was clearly
wrong. We find that the continuum fit does not have a significant
effect on the {\it{shape}} of the resultant mean profile. However the
continuum fit chosen does, of course, have an impact on the level of
the continuum seen in the mean profiles. In general, it was found that
the final continuum fit selected was at too high a level, resulting in
the (subtracted) continuum wings of the mean profiles lying well below
zero. This is straightforward to remedy by iteratively lowering the
level of the continuum fit until the desired zero continuum level is
achieved in the LSD profile. Again, this process was found not to alter
the shape of the mean profile.

It should be noted that since our line list contains normalised line
depths and our spectroscopic data are continuum subtracted, a further
step was needed to prepare the line list for use with our data. Each
line depth in the line list was multiplied by the value of a continuum
fit to a K5V spectral type template at that wavelength. In this way,
the line list now contains a set of lines that have the correct
relative depths and may be used with continuum subtracted data. The
sensitivity of the LSD process to the choice of line list used has
been investigated by carrying out LSD on the same RU Peg spectrum
using a range of spectral type line lists that covers the spectral
types K0V to K8V inclusive. As in \citet{barnes_phd}, there is no
evidence that slight errors in the assumed spectral type, and therefore an
incorrect line list, introduces any appreciable systematic errors
in the final average profile.

\begin{figure*}
\includegraphics[width=16cm]{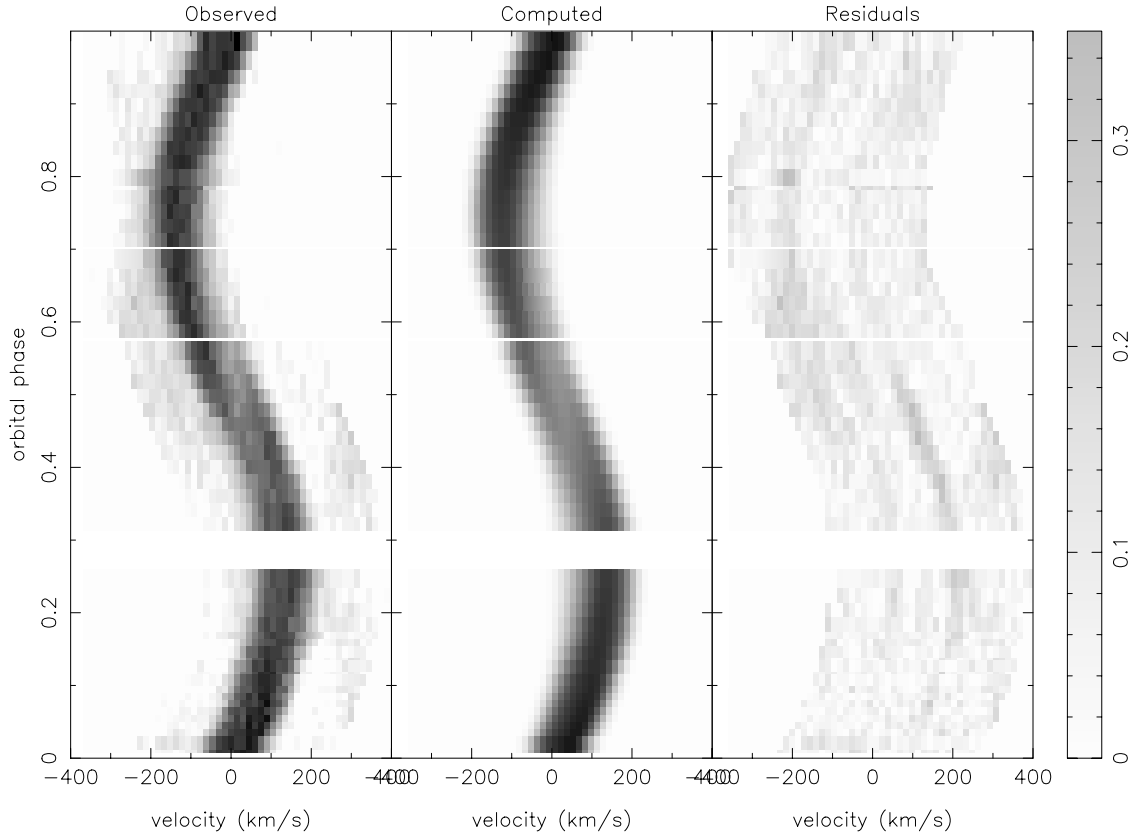}

\caption{{\texttt{Roche}} input trailed spectrum, computed data and
  residuals for RU Peg blue data. Dark grey scales depict deeper line
  depths relative to lighter grey scales.}

\label{trails_blue}
\end{figure*}

\section{Roche tomography}

Roche tomography is an astro-tomographic technique developed to image
the surface of the secondary star in cataclysmic variables
\citep[]{rutten_dhillon_1994,  watson_phd}. The procedure is, by now,
robustly tested and artefacts caused by systematic errors are well
understood (see \citealt{watson_dhillon_2001} for details). Early work
focused on mapping single absorption or emission lines
\citep[see][]{watson_et_al_2003} while, more recently, efforts have
been made to map star-spot patterns (see \citealt{watson_etal_2006}, 2007a).

Roche tomography takes as its input a trail of phase resolved spectral
line profiles, and produces a map of the secondary star in
three-dimensional real space. The key advantage this technique has
over other indirect imaging methods (such as Doppler tomography) is
that it makes use of all the information contained in the line
(i.e. its flux, width and Doppler shift) to map the star. Roche
tomography is, however, very similar to Doppler imaging of single
stars, but for the exceptions that in Roche tomography the star is
Roche-lobe filling, rotates around the binary centre-of-mass, and is
tidally locked. Roche tomography also ignores the
continuum. Successfully applying the technique is often somewhat
harder than Doppler imaging. Roche tomography targets are generally
much fainter than the targets of choice in single star mapping
programs, resulting in spectral profiles of lower signal-to-noise and
poorer quality maps.

We use the tomography code {\it Rochey} developed by \citet{watson_phd} and 
first applied with the addition of LSD by \citet{watson_etal_2006}. 
The synthetic LSD profiles were constructed by assuming that each tile
contributes a Gaussian profile to each line, scaled by projected area and 
intensity. 
Although temperature and abundance will affect the profile, so that the 
spot contribution to the profile will be slightly different from one line 
to another, we don't have enough information to take that into account and
we simply compute an average profile for each tile, and use that average 
profile in 
the mapping. The LSD process is not very sensitive to temperature and 
metallicity effects \citep{barnes_phd}. Although for single stars it is
possible to assume for each tile some combination of `spot' and
`immaculate' assumed spectrum, that is not possible here because irradiation
means that there is no single `immaculate' spectrum -- and indeed the 
contribution of irradiation is something that we want to obtain. We recognise 
that this means that what the {\it Rochey} code produces is essentially a 
contrast map.

\subsection{System parameters}

The use of incorrect system parameters in the Roche tomography
reconstruction results in artefacts being mapped onto the stellar grid
\citep{watson_dhillon_2001}. For example, a set of incorrect values
for the masses and the orbital inclination will result in the code not
being able to properly fit the variations of the profile widths along
the trail. Roche tomography allows a natural method of selecting
optimal values for these parameters, since only the correct values
will give the map of maximum entropy. We carry out reconstructions for
a set of component mass pairs, fitting to the same $\chi^{2}$ each
time, until a full grid of values is obtained -- the so-called
entropy landscape \citep{rutten_dhillon_1994}.

\subsubsection*{Limb Darkening}

An incorrect treatment of the limb darkening for the secondary star
may also result in extra structure being mapped onto its surface. In
this work we make use of a square-root limb darkening law of the
following form \citep{claret_1998},

\begin{equation}
I\left(\mu\right)=I_{0}
\left[1-a_{\lambda}\left(1-\mu\right)-b_{\lambda}\left(1-\sqrt{\mu}\right)\right],
\end{equation}

\noindent where $a_{\lambda}$ and $b_{\lambda}$ are the limb darkening
coefficients at a wavelength $\lambda$. \citet{claret_1998} emphasises
the non-linearity of limb darkening laws applied to model atmospheres
of late K and M type main sequence stars; for the coolest stars, the
departure from linearity is drastic.

Since limb darkening is a function of wavelength, we determine an
effective central wavelength for the blue arm ISIS data of $4875$\AA~
that takes into account the line depths and spectral noise and is
calculated after \citet{watson_etal_2006} using,

\begin{equation}
\lambda_{\rm cen}= \frac{\sum_{\rm i}\left(1/\sigma_{\rm
    i}^{2}\right)d_{\rm i}\lambda_{\rm i}} {\sum_{\rm
    i}\left(1/\sigma_{\rm i}^{2}\right)d_{\rm i}},
\end{equation}

\noindent where $d_{\rm i}$ and $\sigma_{\rm i}$ are the flux and error at
wavelength $\lambda_{\rm i}$, respectively.

We assume a secondary star spectral type of K5V (see section
{\ref{prv}}) and adopt values for the effective temperature of $4557$K and log
surface gravity (in \cms) of 4.6 
(\citealt{gray_1992}). The closest model available in the database of
non-linear limb darkening coefficients provided by
\citet{claret_1998} has an effective temperature of 4660K
and surface gravity of 4.6. We obtained values for the limb darkening
coefficients in the Johnson $B$ and $V$ bands and linearly
interpolated across them to arrive at values corresponding to the calculated
effective central wavelength of our spectra. These coefficients,
$a_{\lambda}=0.8062$ and $b_{\lambda}=0.0538$, are used in all
subsequent reconstructions. For the red arm ISIS data with an
effective central wavelength of $6426$\AA, we obtain coefficients of
$a_{\lambda}=0.482$ and $b_{\lambda}=0.347$ for the Johnson $R$ band
($\lambda_{\rm cen}=6300$\AA). Note that it was not possible to
interpolate values of the coefficients at our central red wavelength,
because of a lack of limb darkening data for photometric bands later
than $R$ in \citet{claret_1998}.

 Note that we have used continuum limb-darkening coefficients. We recognise that
coefficients for the cores of absorption lines are considerably smaller;
however, the precise values for line limb-darkening coefficients depend on
many factors, such as the angle of incidence, and therefore require detailed
and time-consuming calculations (e.g. \citet{collins_1995}). We have not 
attempted these, for three reasons. Firstly, it is the wings of the lines, where
the difference from the continuum is least, that constrain the rotation and
therefore the mass ratio. Secondly, we have tried varying the limb-darkening
coefficients, and even with no limb darkening the maps still show the same
gross features, although the parameters change slightly. Thirdly, 
the coefficients we have chosen give a somewhat lower $\chi^2$ than other 
values we tried, even though the results are not very sensitive to the choice.

\subsubsection*{Systemic velocity, inclination and component masses}

Figs 6(a) and (b) show surfaces of maximum entropy for a set of
inclinations and systemic velocities for the blue and red ISIS data,
respectively. For each systemic velocity/inclination pair an entropy
landscape has been constructed, and the maximum entropy value
extracted. In practice, an inclination value is first selected based
on previous estimates and a series of entropy landscapes are generated
for a range of systemic velocities.

\begin{figure*}
\centering \subfigure[Surface of maximum entropy (blue ISIS data).]  {
    \label{blue_3dparas}
    \includegraphics[scale=0.28]{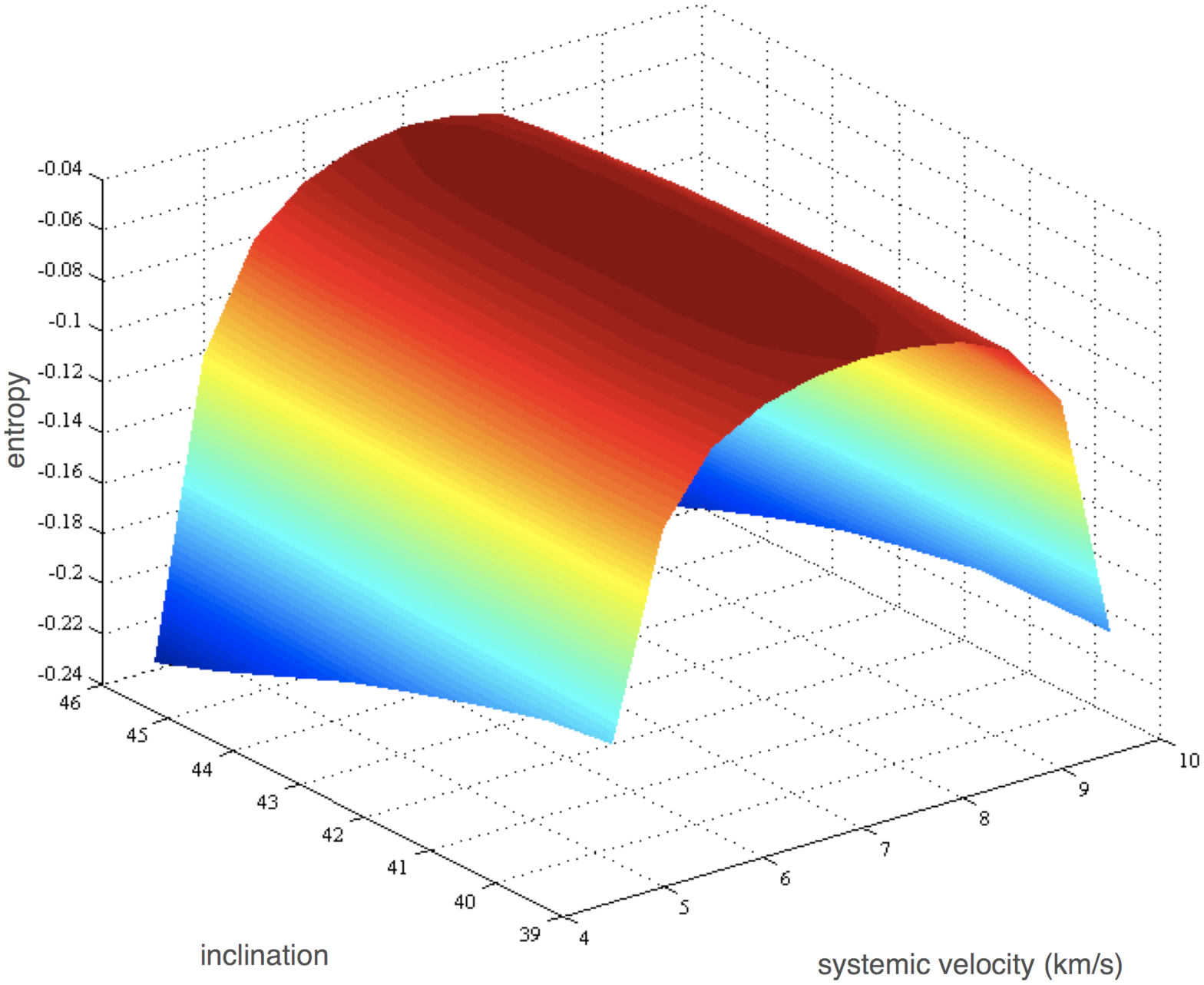} }
\hspace{1.3cm} \subfigure[Surface of maximum entropy (red ISIS data).]
       {
    \label{red_3dparas}
   \includegraphics[scale=0.28]{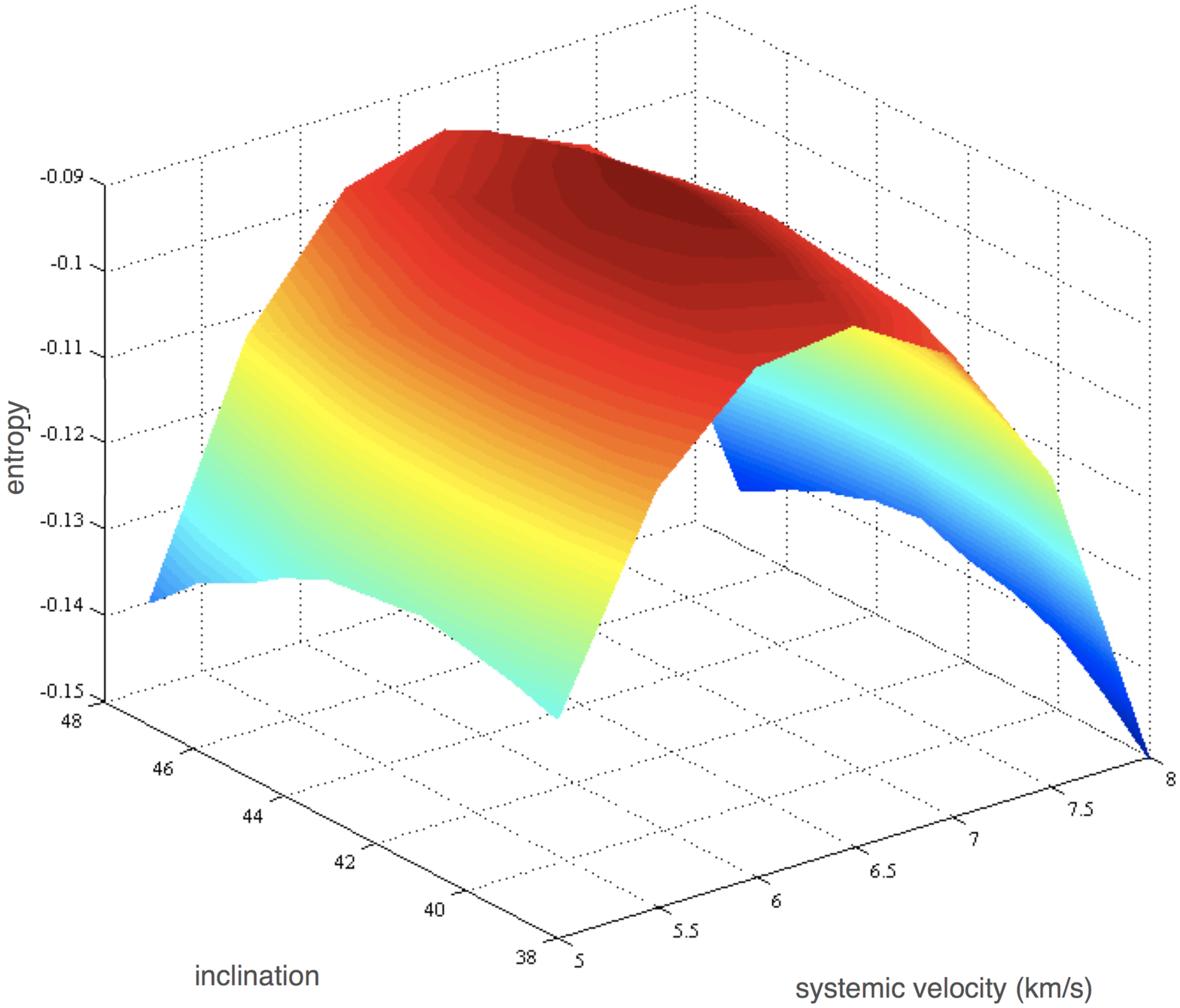} } \subfigure[Maximum
         entropy as a function of systemic velocity (ISIS blue)] {
    \label{rupeg_blue_opt_gamma}
    \includegraphics[scale=0.28,angle=270]{blue_gamma_at43.ps} }
\hspace{1.3cm} \subfigure[Maximum entropy as a function of systemic
  velocity (ISIS red)] {
    \label{rupeg_red_opt_gamma}
    \includegraphics[scale=0.28,angle=270]{red_gamma_at42.ps} }
\subfigure[Maximum entropy as a function of inclination (ISIS blue)] {
    \label{rupeg_blue_opt_incl}
    \includegraphics[scale=0.28,angle=270]{blue_incl_atg7.ps} }
\hspace{1.3cm} \subfigure[Maximum entropy as a function of inclination
  (ISIS red)] {
    \label{rupeg_red_opt_incl}
    \includegraphics[scale=0.28,angle=270]{red_incl_atg6p5.ps} }

\caption{Optimal system parameters. In the top panels, the value of
  each point on the surface is obtained from an entropy landscape
  carried out at a particular combination of orbital inclination and
  systemic velocity. The optimum systemic velocity is clearly
  independent of the assumed inclination. On this scale however, it is
  unclear where the peak inclination lies. The lower panels show
  sections through the surfaces (a) and (b), and allow the peaks to be
  determined (see text).}

\label{singlelinemaps}
\end{figure*}

Previous Roche tomography work (Watson et al. 2003, 2006, 2007a) has
shown that
for almost any choice of inclination (and indeed for a wide range 
of mass pairs) the entropy landscape yields the same systemic velocity.
This is also true here,
where for the blue data, the optimum systemic velocity of $7\kms$ is
constant for the range of inclinations
probed. Fig. {\ref{rupeg_blue_opt_gamma}} shows a slice through the
entropy surface shown in Fig. {\ref{blue_3dparas}} at an inclination
of $43^{\circ}$, showing a peak in the entropy at $7\kms$. Similarly,
Fig. {\ref{rupeg_red_opt_gamma}} shows a cut through the red maximum
entropy surface at an inclination of $42^{\circ}$, where the entropy
clearly peaks at a systemic velocity of $6.5\kms$. It is comforting to
note that the optimal systemic velocity derived from two essentially
independent data sets is constant for the full range of inclinations
probed, 
 and has essentially the same value for the two data sets.

We are unable to assign a fully rigorous (Monte Carlo) error determination 
as such a procedure would be 
vastly computationally expensive (see also below). Test reconstructions, 
using values for the systemic velocity on either side of our optimum value, 
result in rapid degradation of the image quality, 
as also found by Watson et al. (2007a), who reported that ``reconstructions 
were almost impossible for assumed systemic velocities that differed by more 
than $\pm2\kms$ from the optimal value''. 
In addition, we found that viable reconstructions are
 impossible at inclination values $\pm10^{\circ}$ from our optimum value. 
It is therefore unlikely that the errors exceed half these extreme values.

 While our optimal systemic velocity 
is close to the value derived
from the more traditional radial velocity study in Section
\ref{secondary_star}, which yielded 
$10.2\pm0.4\kms$ for the
systemic velocity, it does not agree with it within the error bars. This is
not too surprising, since the value obtained through the radial
velocity curve is subject to an incomplete treatment of the surface
distribution of flux on the secondary, which is known to cause
systematic errors in the systemic velocity determination. Indeed, as
noted earlier, there remain residual radial velocities of up to $\sim$10\%
at some orbital phases after fitting a circular orbit to the
radial velocity data.

\citet{kiplinger_1979} used the earlier radial velocity data of
\citet{kraft_1962} to derive a value of $-1\pm5\kms$ for the systemic
velocity (though he was unable to detect the secondary star in this
work due to severe veiling of the secondary lines by a bright disc on
decline from outburst). It should be noted that the work of
\citet{kraft_1962} represents the earliest detection of RU Peg as a
binary, and the radial velocity data reported there are derived from
very low resolution data, with the subsequent orbital fit showing a
large scatter. Our value is consistent with \citet{stover_1981}, who
reports a systemic velocity of $5\pm4\kms$ derived from observations
of the system taken in quiescence with higher quality
data. \citet{friend_et_al_1990a} quote a slightly larger value of
$13\pm2\kms$ using the NaI doublet, though they question whether or
not the secondary star semi-amplitude is truly accurate, or if it is
biased by low level irradiation. It is noted here that an accurate
value for the radial velocity of the template used in the
cross-correlation is also important, since this must be subtracted
from the CV radial velocity data prior to orbital fitting. In our
case, this has been accurately measured using LSD.

Once the systemic velocity has been determined, it is then possible
for the inclination and the component masses to be determined. These are
solved for essentially simultaneously, although for computational convenience
we in practice chose a value of $i$ and produced an entropy landscape for the
two masses, then repeated this process for a range of $i$, choosing 
finally the inclination and mass pair with the highest overall entropy.

We first show how the entropy depends on the choice of $i$.
Fig. {\ref{rupeg_blue_opt_incl}} shows the maximum entropy values
extracted from a series of mass landscapes obtained over a range of
inclinations with a fixed systemic velocity of $7\kms$ using our blue
data. Similarly in Fig. {\ref{rupeg_red_opt_incl}}, the maximum
entropy is plotted assuming the systemic velocity of $6.5\kms$ derived
earlier, for the red ISIS data. For the blue arm, we obtain a value of
$43^{\circ}$
For the red arm data, this value shifts slightly to $42^{\circ}$;
just as we chose a unique
value of $\gamma$ based upon the relative quality of the mean profiles
obtained from the blue and red data, we also chose to adopt the blue
arm derived value of $43^{\circ}$
for the inclination of RU Peg. 

It is at first sight surprising that an inclination can be deduced for a 
non-eclipsing system. However, the inclination affects the line profile
in two ways. First, the variation in the apparent width of the stellar
absorption line-profile around the orbit depends on $i$
(e.g. \citealt{shahbaz98}); second, the way the spot bumps traverse the
profile depends
quite strongly on $i$ (for example, there would be no motion of the bump at all 
if the inclination were zero). The maximum entropy method disentangles this
dependence and yields an inclination. As with the systemic velocity, we would ideally 
like to estimate all uncertainties rigorously by undertaking a full Monte
Carlo study; however, we have estimated that this might take up to a month 
of computer time, and no-one doing Roche tomography studies has yet
attempted a full error analysis.

\citet{friend_et_al_1990a} find a value of $33^{\circ}\pm5^{\circ}$
for the inclination derived by assuming a main-sequence mass-radius
relation for the secondary. However, if the primary star is a
carbon-oxygen white dwarf (i.e. $M_{1}=0.55$--$1.2$\,M$_{\odot}$
-- \citealt{iben_1985}), then the allowed range of inclinations for
RU Peg is $34^{\circ}$--\,$48^{\circ}$, which our value lies well within. The
assumption that the secondary obeys a main sequence mass radius
relation is also applied in \citet{kiplinger_1979} who quotes a lower
limit of $30^{\circ}$. Our value is, however, in good agreement with
\citet{sion_urban_2002}, who determine a value of $41^{\circ}$ from
model fitting of the disc spectrum.

\begin{figure*}
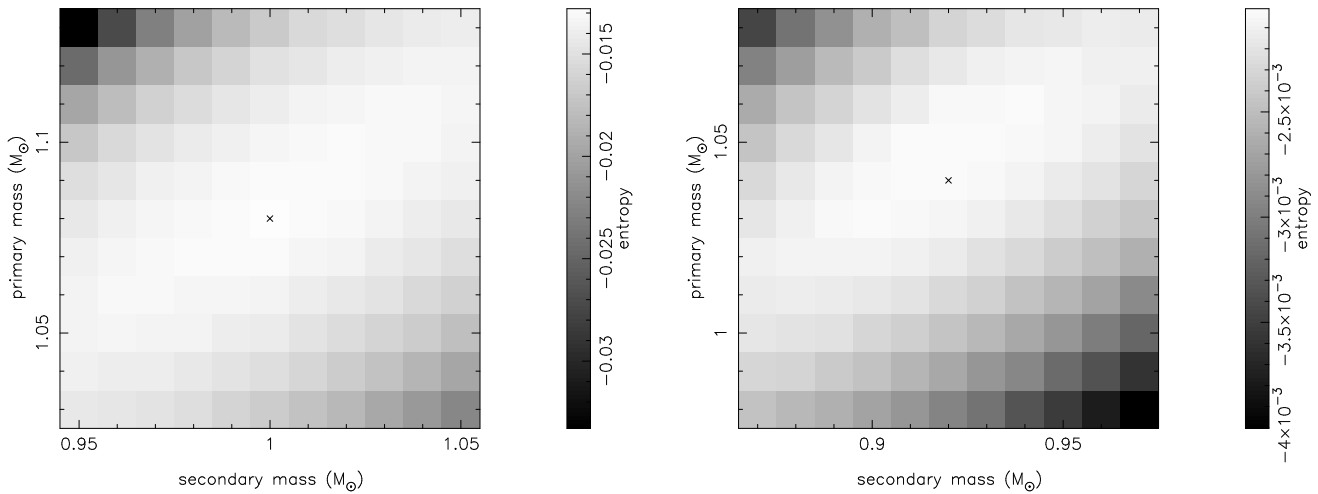

\centering {
    \label{blueland}
    \includegraphics[scale=0.47, angle=270]{blueland.ps} }
\hspace{0.3cm} {
    \label{redland}
    \includegraphics[scale=0.47,angle=270]{redland.ps} }

\caption{{\it{Left}} -- Entropy landscape for the RU Peg blue data,
  assuming an inclination of $43^{\circ}$ and a systemic velocity of
  $7\kms$. Black pixels represent mass combinations for which no
  acceptable solution could be found. The black cross indicates the
  entropy peak, corresponding to primary and secondary masses of
  $M_{1}=1.08$\,M$_{\odot}$ and $M_{2}=1.0$\,M$_{\odot}$. {\it{Right}}
  -- Entropy landscape for the RU Peg red data, assuming an
  inclination of $42^{\circ}$ and a systemic velocity of
  $6.5\kms$. The component masses giving rise to the maximum entropy are
  $M_{1}=1.04$\,M$_{\odot}$ and $M_{2}=0.92$\,M$_{\odot}$.}

\label{landscapes}
\end{figure*}

The mass entropy landscapes corresponding to the above values of systemic
velocity and inclination are shown in Fig. {\ref{landscapes}}.
From these, we derive estimates for the
component masses of $M_{1}=1.08$\,M$_{\odot}$ and
$M_{2}=1.0$\,M$_{\odot}$ for the blue data, and
$M_{1}=1.04$\,M$_{\odot}$ and $M_{2}=0.92$\,M$_{\odot}$ for the red
data. 
While the derived primary star mass values are quite similar,
  there is a greater difference between the values for the secondary
  star masses.  We see no reason to preferentially adopt one estimate
  of the mass (either derived from the blue or the red data) over the
  other. Instead we adopt an average of these values and
  estimated uncertainties of
  $\pm0.04$\,M$_{\odot}$ for the primary masses and
  $\pm0.08$\,M$_{\odot}$ for the secondary masses which take into account the
  scatter between the values, 
but not any possible systematic effects; the real uncertainty is very likely 
to be larger.

Having determined the best-fit system parameters, it is of interest to
see what radial-velocity amplitudes are implied by these values and how
they compare with previous direct determinations. Because the system
parameters from the blue and red data differ, we consider these
data sets separately. The results are given in Table~\ref{semi-amp}, where
they are compared with our direct values from Section~\ref{rv} and values
from the literature.

\begin{table}
\begin{center}
\begin{tabular}{|l|c|c|}
\hline Data & $K_2$ (\kms) & $K_1$ (\kms) \\ \hline 
Red arm, from tomogram & 131.2 & 116.1 \\ 
Blue arm, from tomogram &  133.5 & 123.6 \\ 
Direct, red arm, this paper & 128.2 & 112.6 \\ 
\citet{friend_et_al_1990a} & 121$\pm$2 &  \\ 
\citet{stover_1981} &  & 94$\pm$3 \\ 
\citet{kiplinger_1979} &  & 88$\pm$10 \\ \hline
\end{tabular}

\caption[Comparison of RV amplitudes]{A comparison of the radial
velocity semi-amplitudes implied by our maximum entropy results
with direct measurements.\label{semi-amp}}

\end{center}

\end{table}

The deduced $K_2$ values from the two independent datasets agree satisfactorily
and are not very different from the direct value found in Section~\ref{rv}.
More surprisingly, they are slighter larger than the direct value, which is not
what would be expected if the difference were due to irradiation. This 
probably means that irradiation has no dramatic effect on the radial velocity
amplitude in this long-period, low inclination system, and that the
values are consistent, implying that the uncertainties on each are about
4\kms. While our $K_2$ value of $\sim$130\kms\ differs significantly from
that found by \citet{friend_et_al_1990a}, we note that their 
data was of significantly lower resolution (at best 75\kms) and this may
account for the difference. The deduced $K_1$ 
values are rather more different from one another. If they 
are also assumed to be free from systematic errors, then this would
indicate an uncertainty in the derived $K_1$ values of $\sim$5\kms.
However, the three values from this paper are sufficiently 
similar to rule out the considerably smaller values found by 
\citet{stover_1981} and \citet{kiplinger_1979}.

\subsection{Surface images}

\begin{figure*}
\centering {
    \label{blue_map}
    \includegraphics[scale=0.55]{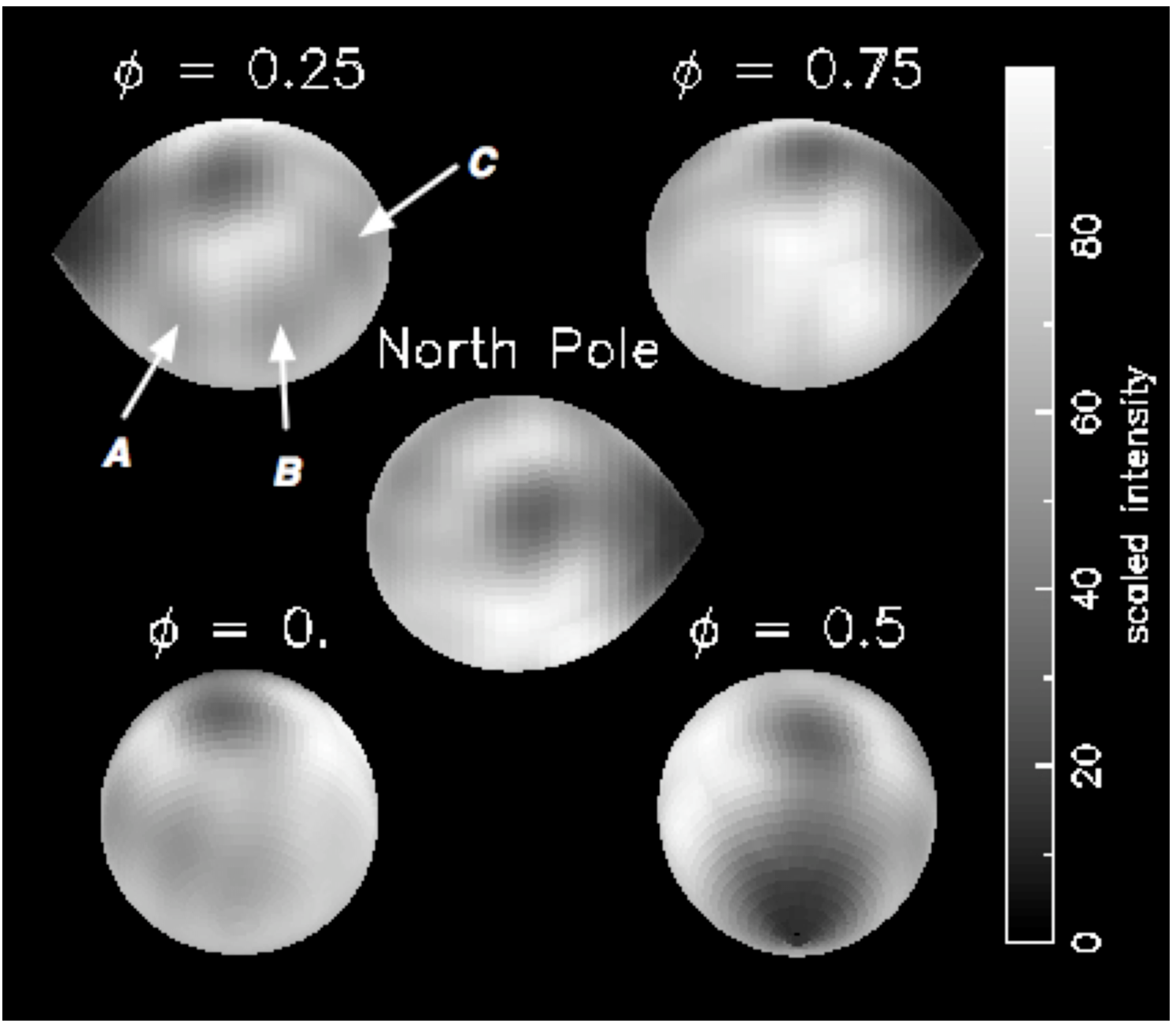} }
\hspace{0.3cm} {
    \label{red_map}
    \includegraphics[scale=0.573]{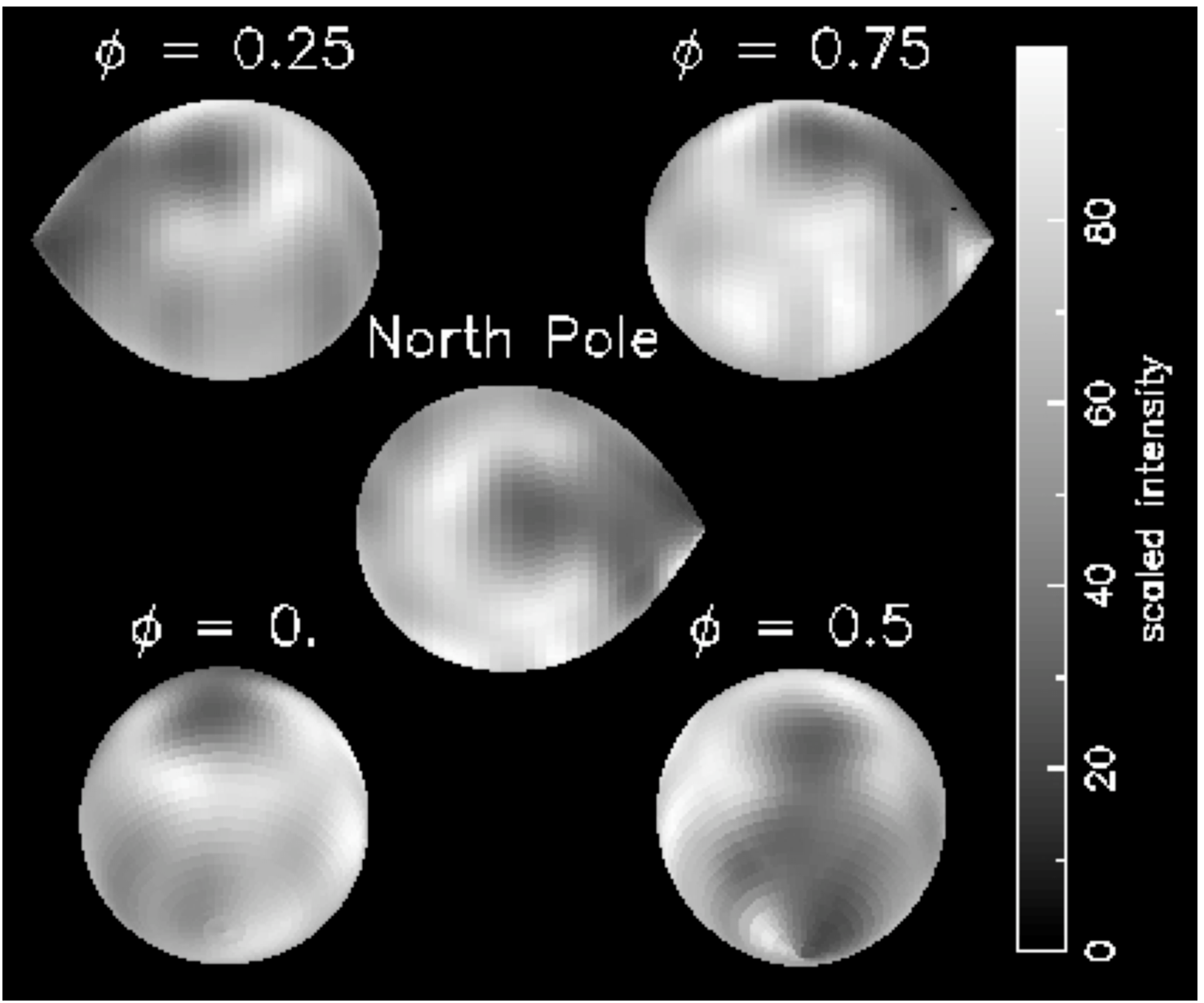} }

\caption{{\it{Left}} -- Roche tomogram of RU Pegasi derived from
  blue-arm ISIS data. The intensity scale is adjusted so that dark
  scales represent regions of lower absorption. The map is presented
  as the observer would see it at an inclination of $43^{\circ}$
  except for the central image, where the observer is directly above
  the north pole. {\it{Right}} -- Roche tomogram of RU Pegasi derived
  from red-arm ISIS data. The grey scale and viewpoint are the same as
  in the blue-data map.}

\label{maps}
\end{figure*}

Fig. {\ref{trails_blue}} shows the input trailed spectrum of the
blue-arm LSD absorption profiles together with the computed data from
Roche tomography and the corresponding residuals. The Roche tomogram
produced from this trail is shown in full in the left hand panel of
Fig. {\ref{maps}}.  Perhaps the most visually striking feature is the
large, dark, high latitude star-spot. This is a surprising result given
the intermediate resolution of the ISIS spectrograph which we had
assumed would be insufficient to map spots. However, given the
inclination of the system (43$^{\circ}$), polar features would be
visible throughout the whole orbit and would contribute at every
phase, thereby lending themselves to detection with even relatively
low resolution instruments.

Similar large high latitude spots are commonly found in Doppler
imaging studies of rapidly rotating single stars and have also been
recently detected in two other CVs -- AE Aqr \citep{watson_etal_2006}
and BV Cen \citepalias{watson_etal_2007a}. These features are in clear
contrast to sun-spots, whose locations are rarely more than
30$^{\circ}$ from the solar equator. In the case of RU Peg, we have
determined that the centre of the spot lies at a latitude of
$\sim$82$^{\circ}$, and covers an area approximately 4 per cent of the
total surface area of the secondary star.  What causes the emergence
of such high latitude star-spots? \citet{Schuessler_1992} propose that
the strong Coriolis forces dominate over the buoyancy of emerging
magnetic flux tubes from deep within the convective zone and force
them to follow a path almost parallel to the rotational axis. Although
this conclusion was drawn for single stars, the dynamics of magnetic
flux emergence is essentially the same for synchronously rotating
stars. An alternative scenario of migrating star-spots has been
suggested by \citet{Vogt_etal_1999}, who report on the apparent
clockwise spiral migration of small long-lived spots towards the pole,
perhaps merging with the polar spot. Whether these migrating spots
reinforce or cancel out the polar spot is of course dependent on their
(as yet) unknown polarity. The star-spot in our map also appears to be
deflected slightly towards the trailing edge of the star. This
apparent deflection is also clear in the Roche tomograms of AE Aqr and
BV Cen by Watson et al. (2006, 2007a). A comparison of the Roche
tomogram of RU Peg to those of AE Aqr and BV Cen is shown in
Fig. {\ref{high_res_maps}} along with a preliminary map of V426 Oph by
\citetalias{watson_etal_2007b}. Is this a common feature of tidally
distorted rapidly rotating stars? If so, what is the exact mechanism
that produces it? Perhaps it is caused by the orbital motion of the
binary. Interestingly, there is no such high-latitude spot present in
the V426 Oph map, though this system was caught in an extended (and
unusual) quiescent state, where its regular outbursting behaviour had
essentially halted for over a year. Thus, if magnetic activity is
responsible for the behaviour of CVs then the map of V426 Oph may not
be representative of the system as it would normally appear.

Given the resolution of our RU Peg data, this is probably the best
that can be produced by applying LSD. The amplitude variation in the
mean line profile due to small scale spots is too low to be
confidently detected. However, Fig. {\ref{maps}} also seems to suggest
the presence of smaller amplitude features at lower latitudes
(labelled $A$, $B$ and $C$). These could possibly be more star-spots
at the limit of detectability (they seem to be hinted at in the red
tomograms too) or they could simply be artefacts of unknown cause. The
other main feature of the map is the dark region around the L$_1$
point, best seen in the image at phase 0.5 in Fig.~\ref{maps}. While
this could be due to a large spot, given the proximity of this region
to a hot irradiating source (particularly prominent if the white dwarf
is especially hot) it is more likely to be due to irradiation. Such
irradiation patterns have been observed in both high and low
resolution maps. The irradiation in RU Peg was previously mapped by
\citet{davey_phd} using sodium flux deficit measurements. The map
produced showed an anomalous region, centred at phase 0.3 on the
trailing hemisphere and extending some way towards the back of the
star. Indeed, this feature extended well beyond the terminator such
that \cite{davey_phd} concluded that it was highly unlikely to be
caused entirely by heating, and instead speculated that it might be
caused by magnetic activity on the secondary star. Our maps show no
sign of this feature, with the accretion pattern being
symmetric. However, the maps of \cite{davey_phd} have no latitude
discrimination, so it is possible that what was being detected was a
superposition of symmetric heating and a high latitude spot that was
located on the trailing hemisphere.

The right hand panel of Fig. {\ref{maps}} shows the Roche tomogram
created using the red-arm data. Again, prominent are the off-axis
polar spot and irradiation pattern, confirming their reality, although
in this map the two features are rather more blended
together. Curiously, there does seem to be some non-uniformity in the
irradiation pattern with less ionisation on the leading hemisphere
side of the L$_1$ point. It is difficult to explain this feature given
the lower signal to noise of the red profiles and is dismissed as most
likely to be an artefact.

\subsubsection*{Simulating the effects of poor relative flux calibration}

Roche tomography maps relative line fluxes, and as such, the input
data are assumed to have an accurate relative flux calibration. In
practice, this is achieved by either obtaining simultaneous
photometric observations, or by observing a comparison star on the
same slit as the main target. In this case we used a comparison star
on the slit (as described in Section~\ref{sec:obs}) to correct for
slit-losses. However, the trailed spectrum of the LSD profiles shown
in Fig. {\ref{trails_blue}} clearly shows some imperfections in the
relative scaling of the parent spectra, despite multiple attempts to
remove this effect during the data reduction stages.

\begin{figure}
\centering \subfigure[AE Aqr] {
    \label{aeaqr_chris}
    \includegraphics[scale=0.195]{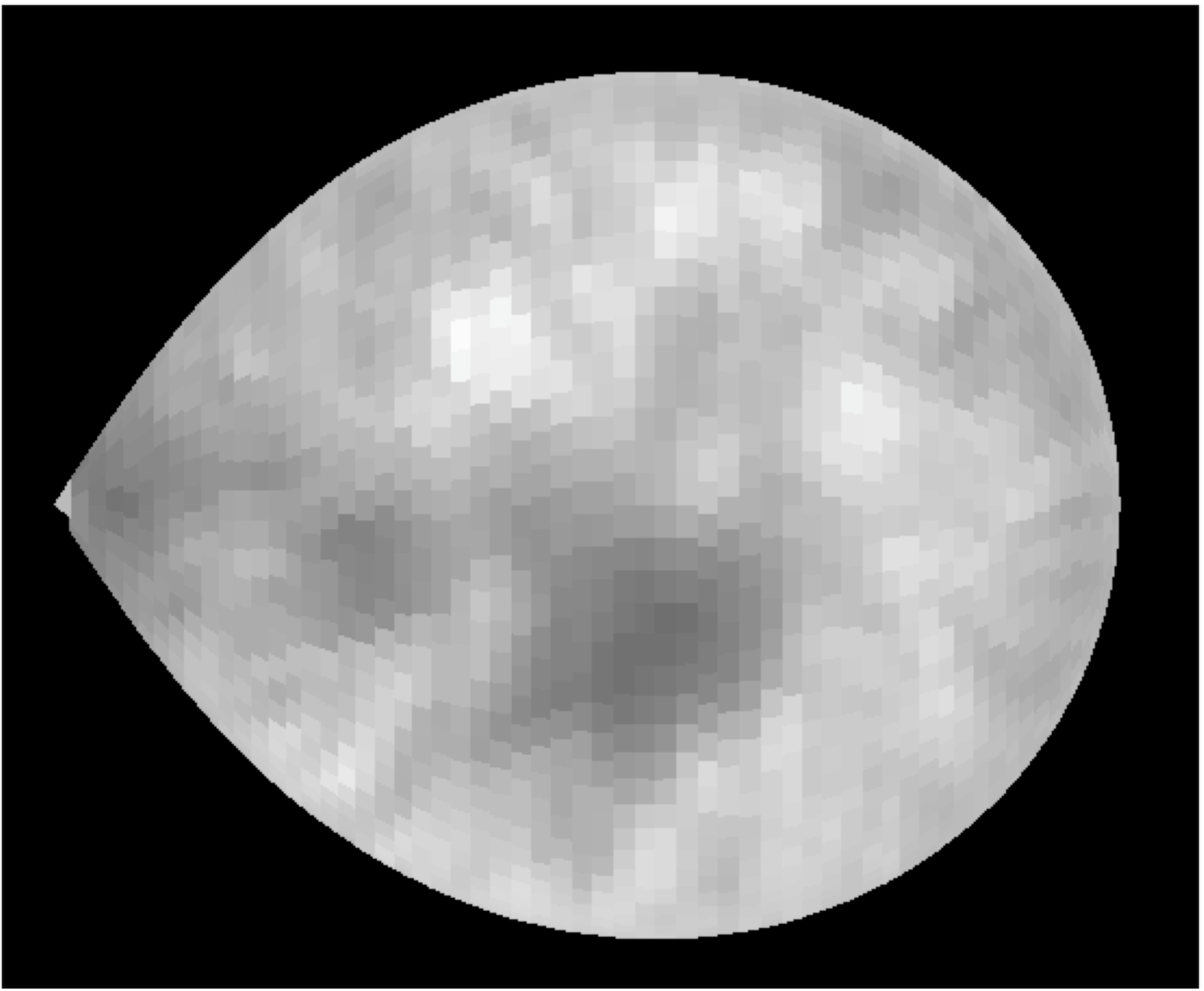} }
\hspace{0.2cm} \subfigure[BV Cen] {
    \label{bvcen_chris}
    \includegraphics[scale=0.195]{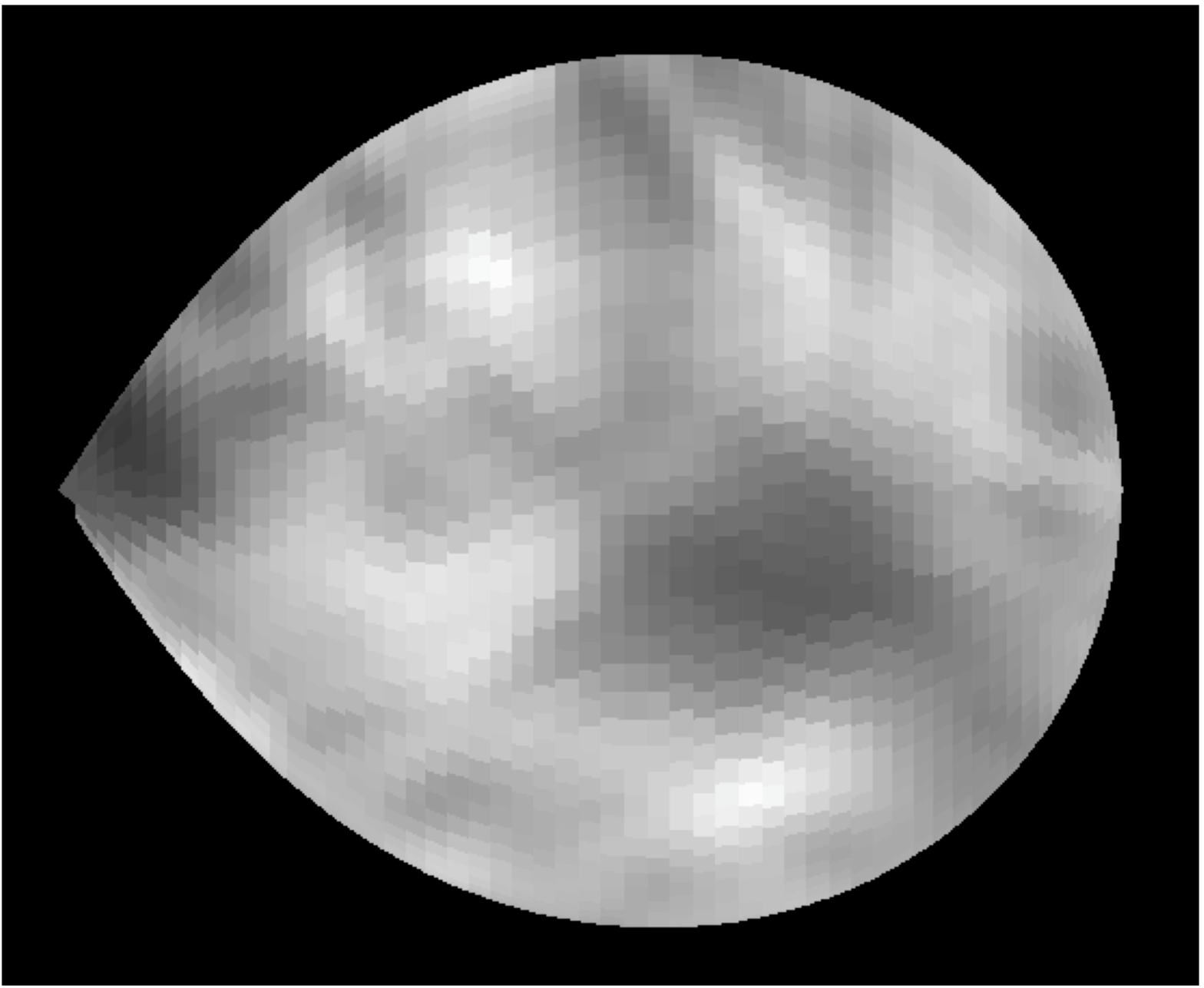} } \subfigure[RU Peg] {
    \label{rupeg_me}
    \includegraphics[scale=0.195]{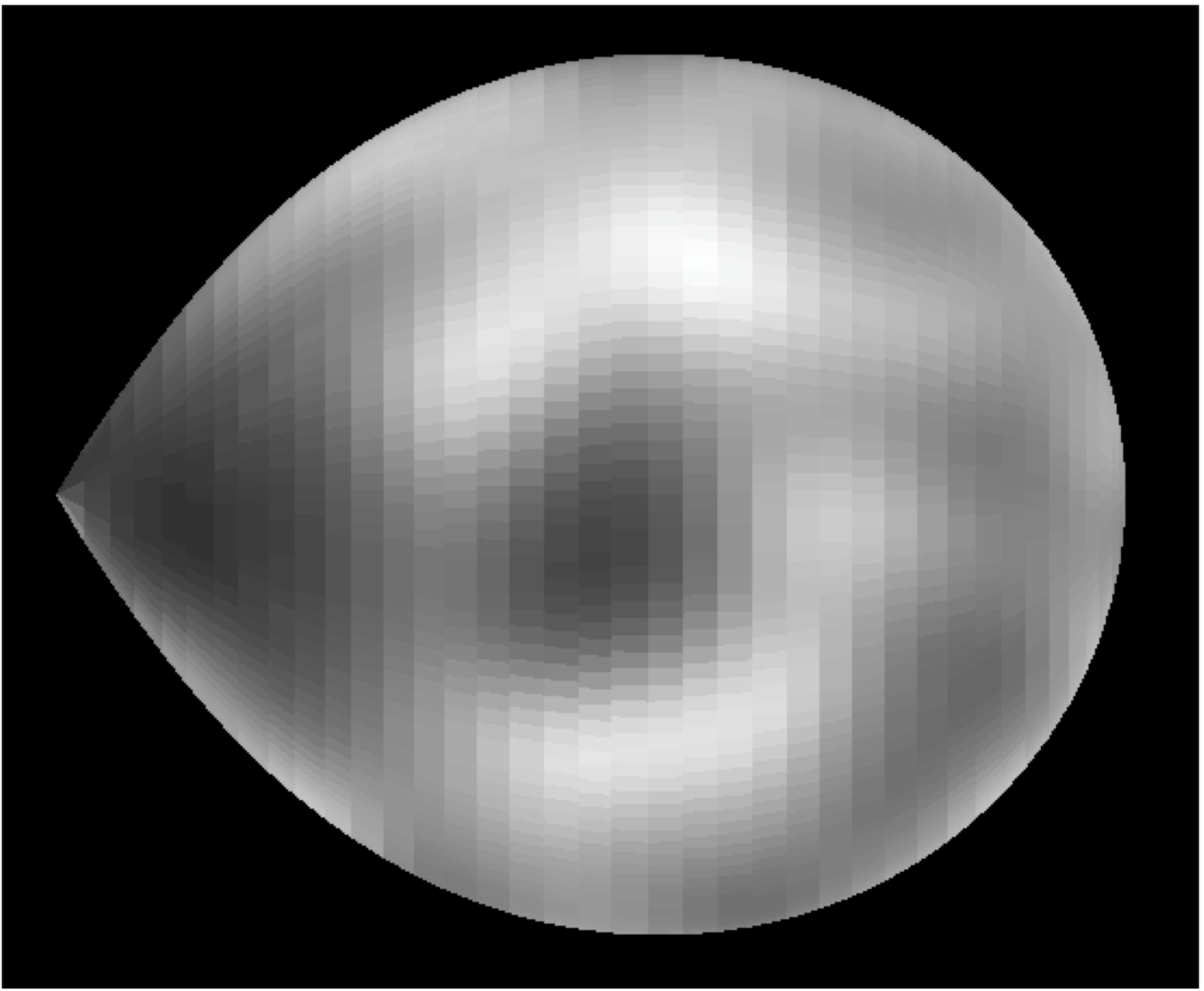} }
\hspace{0.2cm} \subfigure[V426 Oph] {
    \label{v426_chris}
    \includegraphics[scale=0.195]{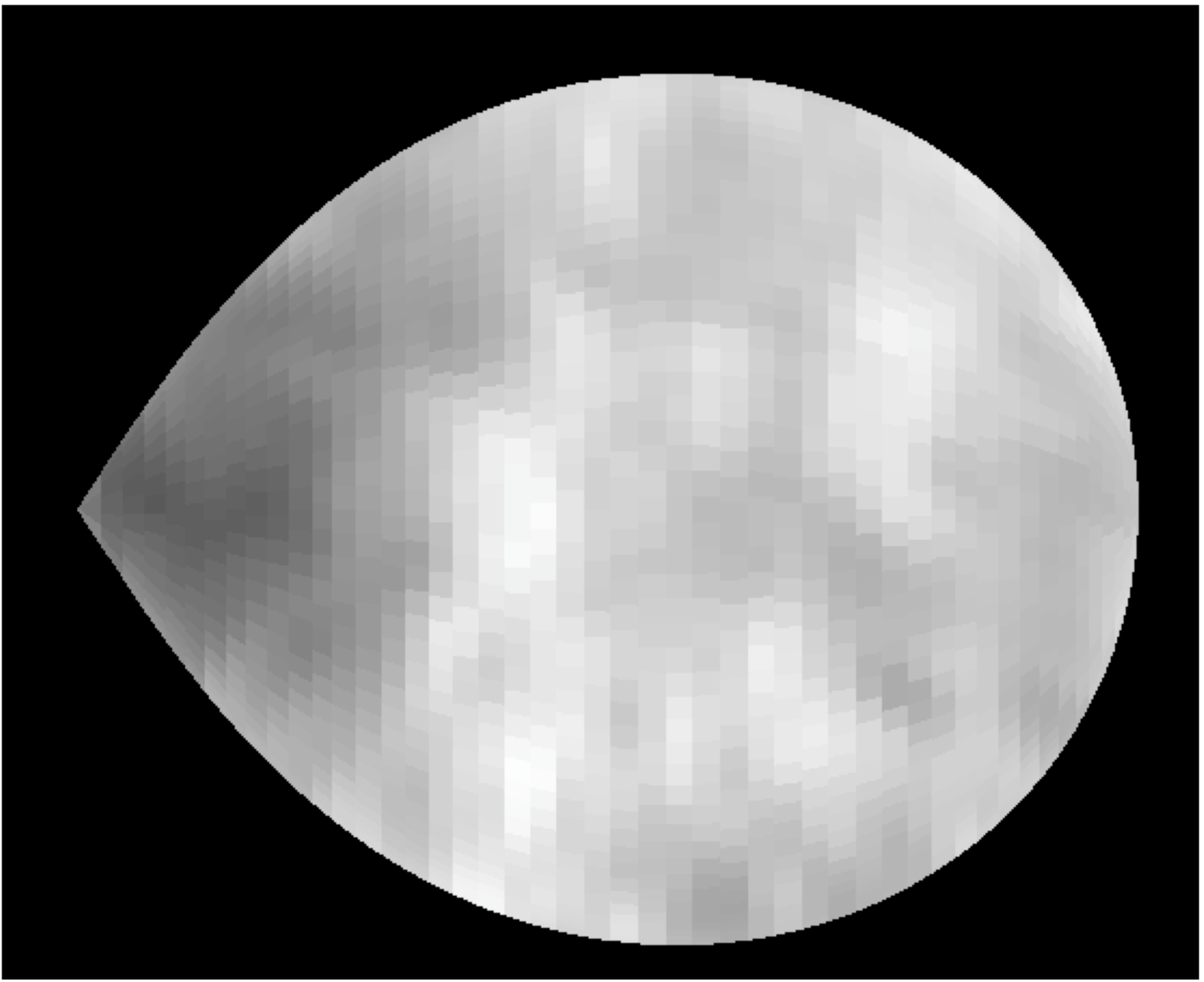} }

\caption{Comparison of published multi-line Roche tomograms with the
  RU Peg map in this work. The maps of AE Aqr and BV Cen both show the
  large high latitude spot present in the RU Peg map while,
  interestingly, the provisional V426 Oph map does not. Note that the
generally smoother appearance of the RU Peg map is caused by the fact that
the data are of lower spectral resolution than for the other three maps, and
so cannot reveal small-scale features; much of the apparent `noise' in the
other maps is probably caused by small spots.}

\label{high_res_maps}
\end{figure}

In order to assess the potential impact this has on the resultant
Roche maps, a simulation has been carried out using a test image
featuring a number of star-spots at a range of longitudes. A very low
noise trailed spectrum was then computed, assuming a negligible
intrinsic profile and an instrumental resolution of
20\kms. Reconstruction of the synthetic trailed spectrum, fitting to a
$\chi^{2}$ of 1.0, showed that all the main features present in the
test image were successfully reconstructed. Next, the trailed spectrum
was given a 10\% error in the relative profile line flux and
reconstructions were once again carried out to the same value of
$\chi^{2}$. Again, the reconstructions faithfully reproduced the
star-spot features, but the stellar surface has an additional noise
pattern superimposed upon it. It is found that this low level noise is
removed by slightly under-fitting the data, to a $\chi^{2}$ of 1.5. 
Therefore, just as
incorrect input parameters have a known effect upon the level of
structure in a Roche map, imperfect relative flux scaling also
produces image artefacts that can be identified. These simulations
give us more confidence that both the irradiation pattern (in general,
we do not believe the asymmetry apparent in the red data)
and the high-latitude spot are real features.
However, the lower resolution of the RU Peg data means that no
smaller-scale features can be reliably identified (cf. 
Fig.~\ref{high_res_maps}).

\begin{figure}
\begin{center}
\hspace{0.5cm}
\includegraphics*[scale=0.55,angle=90]{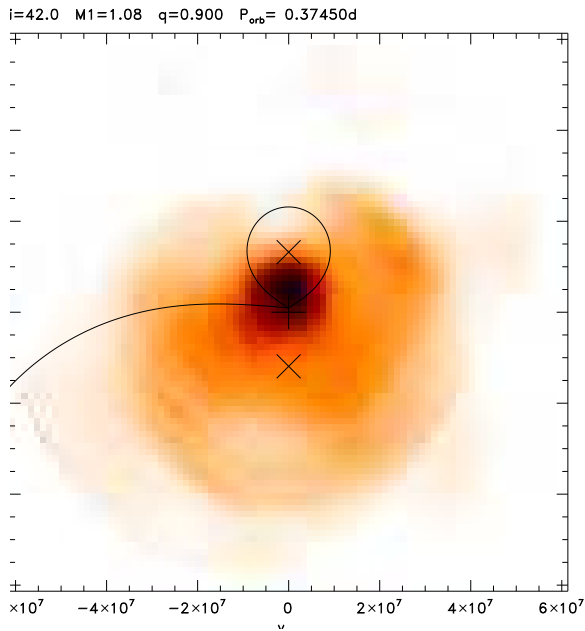}
\end{center}

\caption[Doppler tomogram of RU Peg in the light of H$\alpha$
  emission]{Doppler tomogram of RU Peg in the light of H$\alpha$
  emission. Shown are the centres of mass of the primary and secondary
  stars (crosses) and the surface of the Roche-lobe shaped star. The
  ballistic mass transfer stream is also shown.}

\label{halpha_dopmap}
\end{figure}

\section{H$\alpha$ emission line Doppler tomography}

Doppler tomography is a powerful technique for mapping the emission
lines in CVs. Since our RU Peg data cover most of the orbit during an
outburst, and show prominent Balmer emission lines, it seems logical
to construct a Doppler tomogram. Here, we have used the code of
\citet{spruit_1998} to map the distribution of H$\alpha$ in velocity
space. In this reconstruction, the Roche lobe geometry has been set by
specifying the values for the orbital elements and component masses
derived earlier.

It is immediately obvious that there is no strong
disc emission present in the map; this is probably a result of the system 
being in outburst, and the disc becoming optically thick, so that the lines
switch from emission to absorption.
The most striking emission feature is the very bright region at the
velocity of the secondary star. The brightest pixels are found at a
position exactly where the inner hemisphere of the secondary lies
(also marked on the image). This is in excellent agreement with our
Roche tomograms which show symmetric irradiation about the L$_1$ point.

\section{Conclusions}

The Roche tomograms of RU Peg show prominent irradiation of the inner
hemisphere of the secondary star. The degree of irradiation found is
of no surprise because RU Peg was caught near its peak magnitude
during outburst. Unlike the Na-flux deficit mapping of RU Peg carried out by
\cite{davey_phd}, we see no asymmetry in the irradiation pattern, nor do
we see any evidence for irradiation extending beyond the terminator
towards the back of the star.

The Roche tomograms also show a near-polar spot centred at a
latitude of 82$^{\circ}$ and covering an estimated 4 per cent of the
stellar surface. While the resolution of the data is too low to place
any limit on the overall spot filling factor across the star, the
presence of a large spot comparable to the largest spots seen on BV
Cen and AE Aqr suggests that RU Peg is also highly magnetically
active. As with the prominent high latitude spots found on AE Aqr and
BV Cen, the spot on RU Peg also lies towards the trailing hemisphere.
(Though we note that, at a latitude of 82$^{\circ}$, the spot on
RU Peg is distinctly more `polar' than the prominent spots on
both AE Aqr and BV Cen which lie at a latitude of $\sim$65$^{\circ}$.)
Whether this is a common feature of all CV donor stars and indicative
of a preferred longitude for magnetic flux emergence caused by either
orbital or tidal effects remains to be seen.

Finally, the spot on RU Peg
almost `merges' with the irradiated region. Since spotted
and irradiated regions both appear as apparent absorption-line flux
deficits, we speculate that this may be able to explain the
asymmetric irradiation patterns found by Na-flux deficit mapping of
CV donors (see \citealt{davey_phd}; \citealt{davey92}; \citealt{davey96};
\citealt{catalan99}). Instead of just
indicating irradiated regions, could these maps also be affected
by the presence of large spots which are shifted towards one hemisphere?
With the comparatively low-resolution afforded by these maps, it is feasible
that large spotty regions could become blended with the irradiated
zone. This inability to distinguish between large spots and irradiated
regions in such methods could lead to maps suggestive of an
asymmetric irradiation pattern which may appear to extend beyond the
terminator depending on the location of the spot. This explanation
has the advantage of being able, in principle, to explain asymmetries of
either sign unlike previous explanations (see \citealt{smith95} for a
review of the early maps and their interpretations).

\section*{Acknowledgments}

Alex Dunford was supported by a PPARC studentship. The William
Herschel Telescope is operated by the Isaac Newton Group at the
Observatorio del Roque de los Muchachos of the Instituto de
Astrof\'{\i}sica de Canarias. The authors thank Tom Marsh for the use
of his \texttt{PAMELA} and \texttt{MOLLY} software packages. We
acknowledge the use of the Vienna Atomic Line Database (VALD) for
obtaining our atomic line lists. Finally, we thank Roger Pickard,
Director of the BAA VSS, for arranging amateur photometric
observations that covered the period of our observing run.

\bibliographystyle{mn2e}
\bibliography{references}

\bsp

\label{lastpage}

\end{document}